\documentclass{aa}\usepackage{graphics,epsf}
\newcommand{\lta}{\;
  \raise0.3ex\hbox{$<$\kern-0.75em\raise-1.1ex\hbox{$\sim$
  }}\;\hskip-2pt }
\newcommand{\gta}{\;
  \raise0.3ex\hbox{$>$\kern-0.75em\raise-1.1ex\hbox{$\sim$
  }}\;\hskip-2pt }
\begin{document}
\thesaurus{06(06.13.1; 02.03.1; 02.13.1)}
\title{Spatiotemporal fragmentation and the uncertainties in the solar rotation law}
\author{Eurico Covas\thanks{e-mail: eoc@maths.qmw.ac.uk}\inst{1}
\and Reza Tavakol\thanks{e-mail: reza@maths.qmw.ac.uk}\inst{1}
\and Sergei Vorontsov\thanks{e-mail: svv@maths.qmw.ac.uk}\inst{1,2}
\and David Moss\thanks{e-mail: moss@ma.man.ac.uk}\inst{3}
}
\institute{Astronomy Unit, School of Mathematical Sciences,
Queen Mary, University of London, Mile End Road, London E1 4NS, UK
\and Institute of Physics of the Earth, B.Gruzinskaya 10, Moscow 123810, Russia
\and Department of Mathematics, The University, Manchester M13 9PL, UK
}
\offprints{\em E.\ Covas}
\authorrunning{Covas {\em et al.}}
\titlerunning{Spatiotemporal fragmentation and the solar rotation law}
\date{\today}
\maketitle
\begin{abstract}
Analyses of recent helioseismic
data
indicate that the dynamical regimes at the base
of the convection zone can be different from those
observed at the top, having either significantly shorter
periods or non--periodic behaviour.

Recently spatiotemporal fragmentation/bifurcation has been
proposed as a dynamical mechanism to
account for the multi-mode
behaviour that is possibly observed in the
solar convection zone, without requiring
separate physical mechanisms with different time scales at
different depths.

Here we study the robustness
of this mechanism with respect to changes to
the zero order rotation profile,
motivated by the uncertainties of and
differences between the various reductions of the
helioseimic data.
We find that spatiotemporal fragmentation
is a common feature of the reductions investigated.
\end{abstract}


\section{Introduction}
Recent analyses of the helioseismic data, from both the
Michelson
Doppler Imager (MDI) instrument on board the SOHO
spacecraft (Toomre et al.\ 2000) and the Global Oscillation Network Group (GONG)
project (Antia \& Basu 2000),
have provided strong evidence that
the previously observed
torsional oscillations (e.g.\ Howard \& LaBonte 1980; Snodgrass, Howard \& Webster 1985;
Kosovichev \& Schou 1997; Schou et al.\ 1998), with
periods
of about 11 years,
penetrate into the
convection zone
to depths of at least 10 percent in radius.

Further studies of these data have produced interesting, but
rather inconsistent results.
In particular, Howe et al.\ (2000b)
find evidence for the presence of
similar oscillations near the tachocline situated close to
the bottom of the convection
zone, but with a markedly shorter period of about $1.3$ years, whereas
Antia \& Basu (2000) do not find such oscillations (whilst allowing the possibility 
that more irregular variations may be present).
Whatever the true dynamical behaviour at these lower levels turns out
to be, both these results indicate that
the variations in the differential rotation
can have different dynamical behaviour
at the top and the bottom of the solar
convection zone, with the oscillations at
the bottom having either significantly shorter
periods or non--periodic behaviour.

Recently,
{\it spatiotemporal fragmentation/bifurcation}
has been
proposed as a dynamical mechanism to
account for the possible multi-mode
behaviour in different parts of the
solar convection zone (Covas et al.\ 2000a, hereafter CTM1).
Evidence for this mechanism was found
in the context of
a two dimensional axisymmetric mean field
dynamo model in a spherical shell,
with a semi--open outer boundary condition,
in which the only nonlinearity  is the action
of the azimuthal component of the Lorentz force of the
dynamo generated magnetic field on the solar
angular velocity. The
underlying zero order angular velocity was
chosen to be consistent with
the recent helioseismic data.

Our initial study (CTM1) suffered from two major
shortcomings.
The dynamo model employed inevitably contained major approximations,
and the zeroth order rotation profile
used is bound to include uncertainties, given the
uncertainties in the inversion techniques,
as well as the short extent in time over which data sets are so far available.

To address the first shortcoming,
an extensive study was made of the persistence of the fragmentation
following plausible changes in the
details of the dynamo model. This study showed that
the mechanism remains surprisingly robust
in presence of a variety of rather severe changes to
the model (Covas et al. 2001, hereafter CTM2).

Regarding the second shortcoming, CTM2 also included
a preliminary study, allowing the
zero order rotational profile to be given by either
the inversion of the MDI data reported by Toomre et al (2000)
or of the corresponding GONG data (Howe et al 2000b).
This also showed the mechanism to be robust
with respect to such changes to the rotation profile.

Given the significant uncertainties that still remain in the
helioseismic measurements, especially given the
limited length of the data so far available,
it is crucial to study
the robustness of
the spatiotemporal fragmentation mechanism
to a wider range of the rotation profiles
consistent with the observations.
\section{Zero order rotation profile}
New measurements of the solar internal rotation with SOHO MDI data
have been reported recently by Vorontsov et al (2001),
using novel inversion techniques that are
quite different from those employed earlier.
Results of a 2D inversion confirmed the previous findings
that the torsional oscillations (migrating zonal flows)
penetrate deep into the convection zone. However, given the limited accuracy 
of the helioseismic data currently available, they did not allow
the resolution of these oscillations all the way down to the bottom
of the convection zone.
The so-called 1.5D inversion procedure has also been used,
which consists in measuring separately the consecutive terms $\Omega_i(r)$ in the expansion

\begin{equation}
\label{expansion}
\Omega (r, \theta) = \sum_{i=0}^{\infty}
\Omega_{2i+1} (r) \frac{dP_{2i+1} (\cos \theta)}{d (\cos \theta)}
\end{equation}
where $P_{2i+1}$ is the $(2i+1)$th Legendre polynomial,
$r$ is the distance from the centre and
$\theta$ is the co--latitude.

The 1.5D inversion of Vorontsov et al (2001) was principally targeted
at improvement of the spatial resolution of the rotation profile
near and just below the base of the convection zone,
in the tachocline region---the region where the dynamo mechanism
is most sensitive to the rotation profile,
since the largest gradients of the angular velocity appear to
be localized there.

An important feature of the solar internal rotation as seen in the helioseismic
measurements is that the rapid spatial variation of the angular velocity
in the tachocline is governed principally by the first two or three terms
in the expansion (1); the contribution of higher terms appears to be an order
of magnitude smaller, being essentially undetectable in the helioseismic data
available at present. Another important feature is that this dominant part
of the variation of the angular velocity in the tachocline with depth
and latitude, captured by the first few terms in the expansion (1),
does not show any changes with solar activity (Vorontsov et al 2001).
Taken together, these two features suggest considering the torsional oscillations
as a small-amplitude, rapidly-varying (in $\theta$), time-dependent component
of the solar rotation, imposed on a large-amplitude, slowly-varying (in $\theta$),
nearly-stationary component described by the first terms of the expansion (1).

These considerations show that at least a dominant part of
the time-independent (background) component
of solar rotation can be evaluated by a simple truncation of the expansion (1)
to the first two or three terms, making the truncated expansion
a natural candidate for the zero-order rotation profile in the dynamo modelling.
This choice might be more consistent than
using a time average of the inverted rotation
profiles (as in CTM2), 
since the observational data used in the inversions do not
yet cover enough (ca. 5 yr) of a complete solar cycle to allow proper averaging. 
With the zero-order profile derived from the time-independent part of the data
we can also eliminate any possibility that torsional
oscillations produced by the dynamo model originate from the time varying zonal flows
already present (in time-averaged form) in the background rotation profile.

\begin{figure}[!htb]
\centerline{\def\epsfsize#1#2{0.45#1}\epsffile{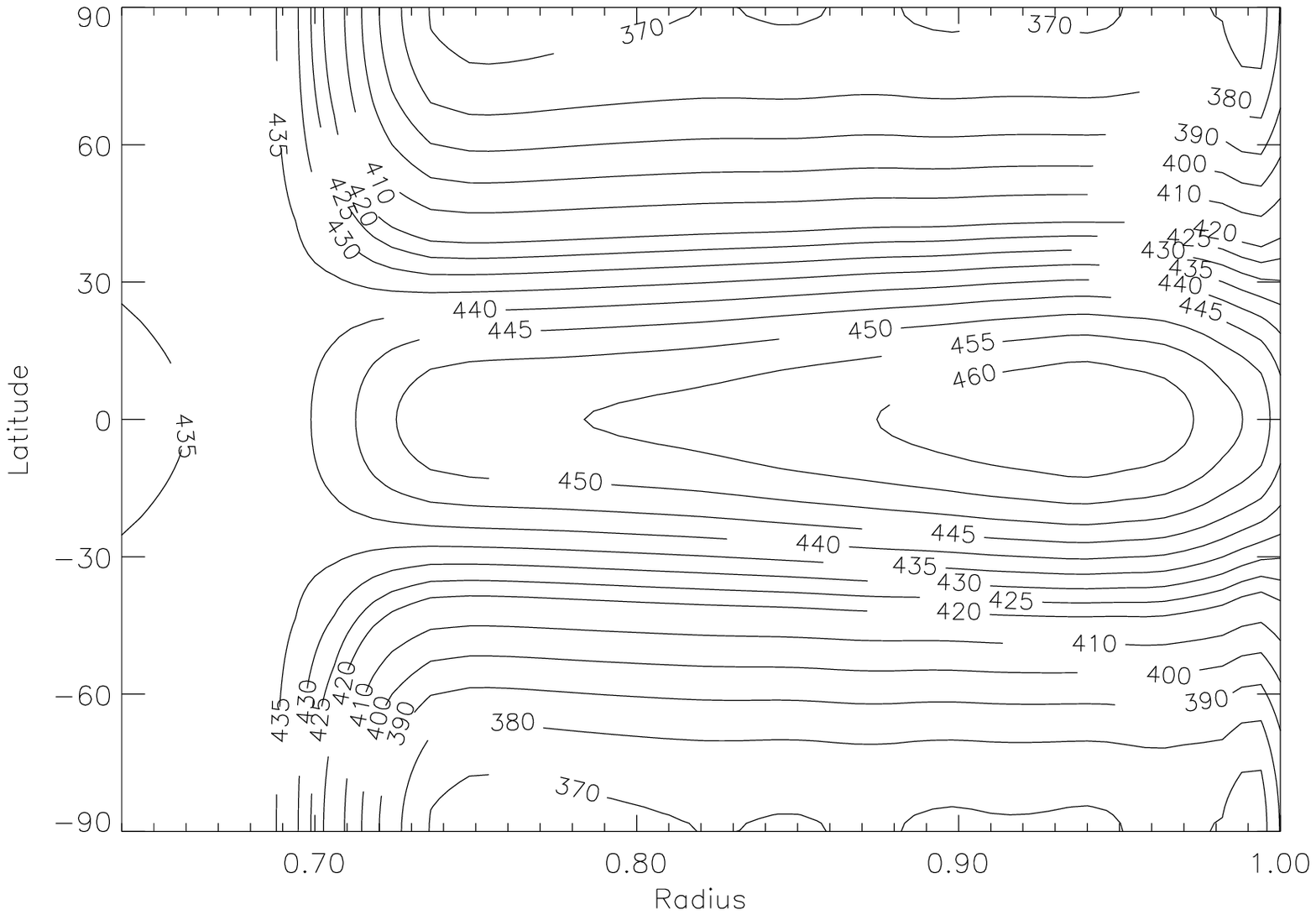}}
\vspace{0.2cm}
\centerline{\def\epsfsize#1#2{0.45#1}\epsffile{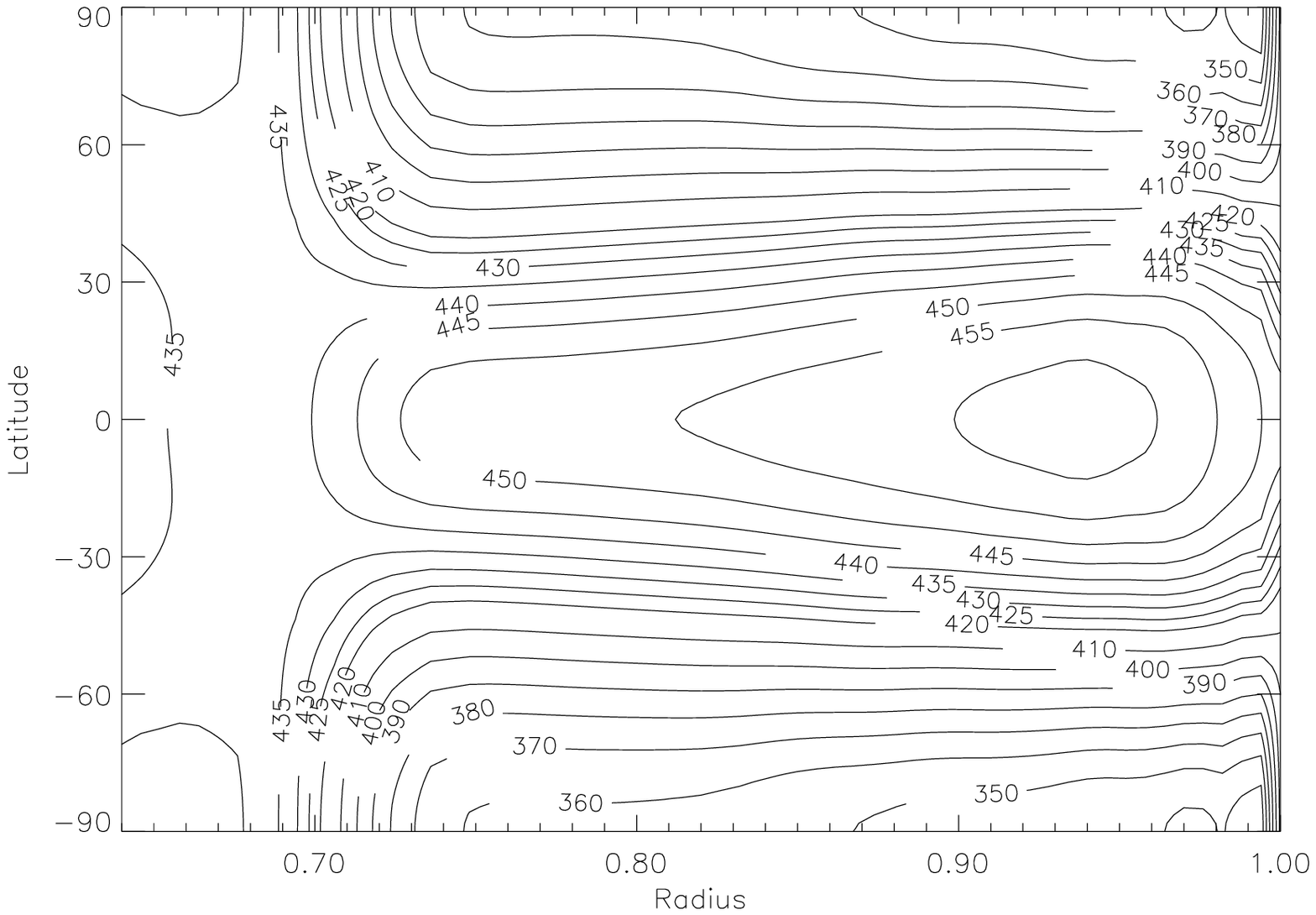}}
\caption{\label{vorontsov135} 
Isolines of the angular velocity of the
solar rotation, obtained from the 1.5-D inversion
of the 360d SOHO MDI data set by Vorontsov et al (2001), given by the first two terms (top panel)
and the first three terms (bottom panel) of the expansion (\ref{expansion}).
Contours are labelled in units of nHz.}
\end{figure}

Thus we here describe a comparative study
of the effects of employing zero order rotation profiles,
obtained by
taking the first two and the first three
terms of the expansion (\ref{expansion}).
These profiles,
inferred from  the 360d SOHO MDI data set,
are shown in Fig. \ref{vorontsov135}.

\section{The model}
In order to test the sensitivity of
the spatiotemporal fragmentation with respect to the 
uncertainties in the rotation profile,
we used the following  dynamo model, which was also
used in CTM1,2.
We assume that the gross features of the
large scale solar magnetic field
can be described by a mean field dynamo
model, with the standard equation
\begin{equation}
\frac{\partial{\bf B}}{\partial t}=\nabla\times({\bf u}\times {\bf B}+\alpha{\bf
B}-\eta\nabla\times{\bf B}).
\label{mfe}
\end{equation}
Here ${\bf u}=v\mathbf{\hat\phi}-\frac{1}{2}\nabla\eta$,
the term proportional to
$\nabla\eta$ represents the effects of turbulent diamagnetism,
and the velocity field is taken to be of the form $
v=v_0+v'$,
where $v_0=\Omega_0 r \sin\theta$, $\Omega_0$ is a prescribed
underlying rotation law and the component $v'$ satisfies
\begin{equation}
\frac{\partial v'}{\partial t}=\frac{(\nabla\times{\bf B})\times{\bf B}}{\mu_0\rho
} . \mathbf{\hat {\bf \phi}}  + \nu D^2 v',
\label{NS}
\end{equation}
where $D^2$ is the operator
$\frac{\partial^2}{\partial r^2}+\frac{2}{r}\frac{\partial}{\partial r}+\frac{1}
{r^2\sin\theta}(\frac{\partial}{\partial\theta}(\sin\theta\frac{\partial}{\partial
\theta})-\frac{1}{\sin\theta})$  and $\mu_0$ is the induction constant.
The sole nonlinearity in the dynamo equation
is the feedback of the azimuthal component of the
Lorentz force (Eq.\ (\ref{NS})), which modifies only slightly the underlying imposed rotation law,
but nevertheless limits the magnetic fields at finite amplitude.
The assumption of axisymmetry allows the field ${\bf B}$ to be split simply
into toroidal and poloidal parts,
${\bf B}={\bf B}_T+{\bf B}_P = B\hat\phi +\nabla\times A\hat\phi$,
and Eq.\ (\ref{mfe}) then yields two scalar equations for $A$ and $B$.
Nondimensionalizing in terms of the solar radius $R$ and time $R^2/\eta_0$,
where $\eta_0$ is the maximum value of $\eta$, and
putting $\Omega=\Omega^*\tilde\Omega$, $\alpha=\alpha_0\tilde\alpha$,
$\eta=\eta_0\tilde\eta$, ${\bf B}=B_0\tilde{\bf B}$ and $v'= \Omega^* R\tilde v'$,
results in a system of equations for $A,B$ and $v'$. The
dynamo parameters are the
two magnetic Reynolds numbers $R_\alpha=\alpha_0R/\eta_0$ and
$R_\omega=\Omega^*R^2/\eta_0$, and the turbulent Prandtl number
$P_r=\nu_0/\eta_0$.
$\Omega^*$ is the solar surface equatorial angular velocity and
$\tilde\eta=\eta/\eta_0$.
Thus $\nu_0$ and $\eta_0$ are the turbulent magnetic
diffusivity and viscosity respectively,
$R_\omega$ is fixed when $\eta_0$ is determined (see Sect.\ \ref{res}),
but the value of $R_\alpha$ is more uncertain.
The density $\rho$ is assumed to be uniform.

Given the ill--determined nature of boundary conditions 
in astrophysical settings, we make physically motivated choices
following CTM1,2.
At inner boundary we chose $B=0$, ensuring angular
momentum conservation, and an overshoot--type condition on ${\bf B}_P$
(cf.\ Moss \& Brooke 2000).
At the outer boundary, we used an open
boundary condition $\partial B/\partial r = 0$ on $B$ and used
vacuum boundary conditions for ${\bf B}_P$
(see  Kitchatinov et al.\ (2000) and CTM1,2 for motivation).

Equations (\ref{mfe}) and (\ref{NS}) were solved using the code
described in Moss \& Brooke (2000) (where more details are given; see also
Covas et al.\ 2000a) together with the above boundary conditions,
over the range $r_0\leq r\leq1$, $0\leq\theta\leq \pi$.
We set  $r_0=0.64$; with
the solar convection zone proper being thought to occupy the region $r \gta 0.7$,
the region $r_0 \leq r \lta 0.7$ can be thought of as an overshoot
region/tachocline.
We note that there is some evidence that the tachocline may be
rather narrower than can be directly resolved by the usual 
inversion techniques. The uncertainties associated with this
possibility are difficult to assess, but we do note that the appearance
of fragmentation appears to be quite robust with respect to changes in
the model (CTM2).
In the following simulations we used a mesh resolution of $61 \times 101$
points, uniformly distributed in radius and latitude respectively.

In this investigation, we took the zero order rotation profile
$\Omega (r, \theta)$ in the region
$0.64\leq r \leq 1$ to be given successively by
the first two and the first three terms of the
series (\ref{expansion}) given above.

For the alpha-effect,  we took $\tilde\alpha=\alpha_r(r)f(\theta)$,
where
$f(\theta)=\sin^4\theta\cos\theta$
(cf.\ R\"udiger \& Brandenburg 1995)
and
$\alpha_r=1$ for $0.7 \leq r \leq 0.8$
with cubic interpolation to zero at $r=r_0$ and $r=1$,
with the convention that $\alpha_r>0$
and $R_\alpha < 0$. Also, in
order to take into account the
likely decrease in the turbulent diffusion coefficient $\eta$
in the overshoot region, we allowed a simple
linear decrease from $\tilde\eta=1$ at $r=0.8$
to $\tilde\eta=0.5$ in $r<0.7$.
We note, however, that as was shown in CTM2, these latter details
of the model are unlikely to remove the possibility of
spatiotemporal fragmentations.
\section{Results}
\label{res}
We calibrated our model so that near marginal excitation
the cycle period was about 22 years.
This determined $R_\omega=44000$, corresponding to
$\eta_0\approx 3.4 \times 10^{11}$ cm$^2$ sec$^{-1}$,
given the known values of $\Omega^*$ and $R$.
With the rotation law given by the first three
terms of (1), the first solutions to be excited in the linear theory
are limit cycles with odd (dipolar) parity with respect to the equator, with
marginal dynamo number
$R_\alpha \approx -3.76$. The even parity (quadrupolar) solutions
are also excited at a similar marginal
dynamo number of  $R_\alpha \approx -3.95$.
These values did not change significantly
when the first two terms of expansion (1) were used.
It is arguable that the turbulent Prandtl number is of the order of unity,
and we set $P_r=1$.
For the parameter range that we investigated, the even parity solutions
can be nonlinearly stable. Given that the Sun is observed to be close to
an odd (dipolar) parity state, and that previous experience shows that small
changes in the physical model can cause a change between odd and even parities
in the stable nonlinear solution, we chose to impose dipolar parity on our
solutions.

With these parameter values,
each of the 
rotation profiles considered here
produced butterfly diagrams which are in qualitative agreement
with the observations.
The model also successfully produced torsional oscillations
that penetrate into the convection zone, in all cases
considered, similar to those deduced from
recent helioseismic data (Howe et al.\ 2000a, Vorontsov et al. 2001), 
and studied in
Covas  et al.\ (2000b)

\begin{figure}[!htb]
\begin{center}
\centerline{\def\epsfsize#1#2{0.42#1}\epsffile{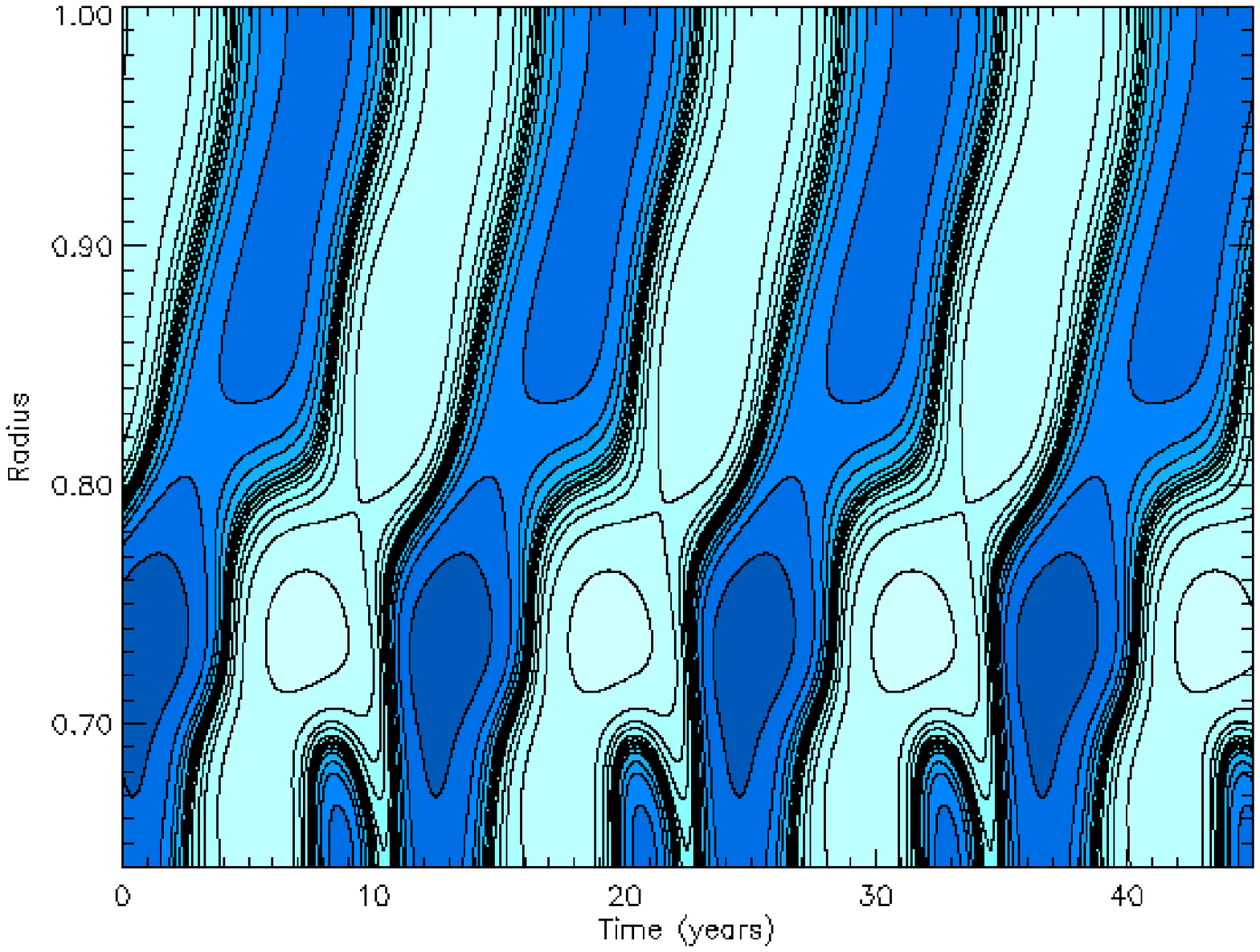}}
\vspace{0.1cm}
\centerline{\def\epsfsize#1#2{0.42#1}\epsffile{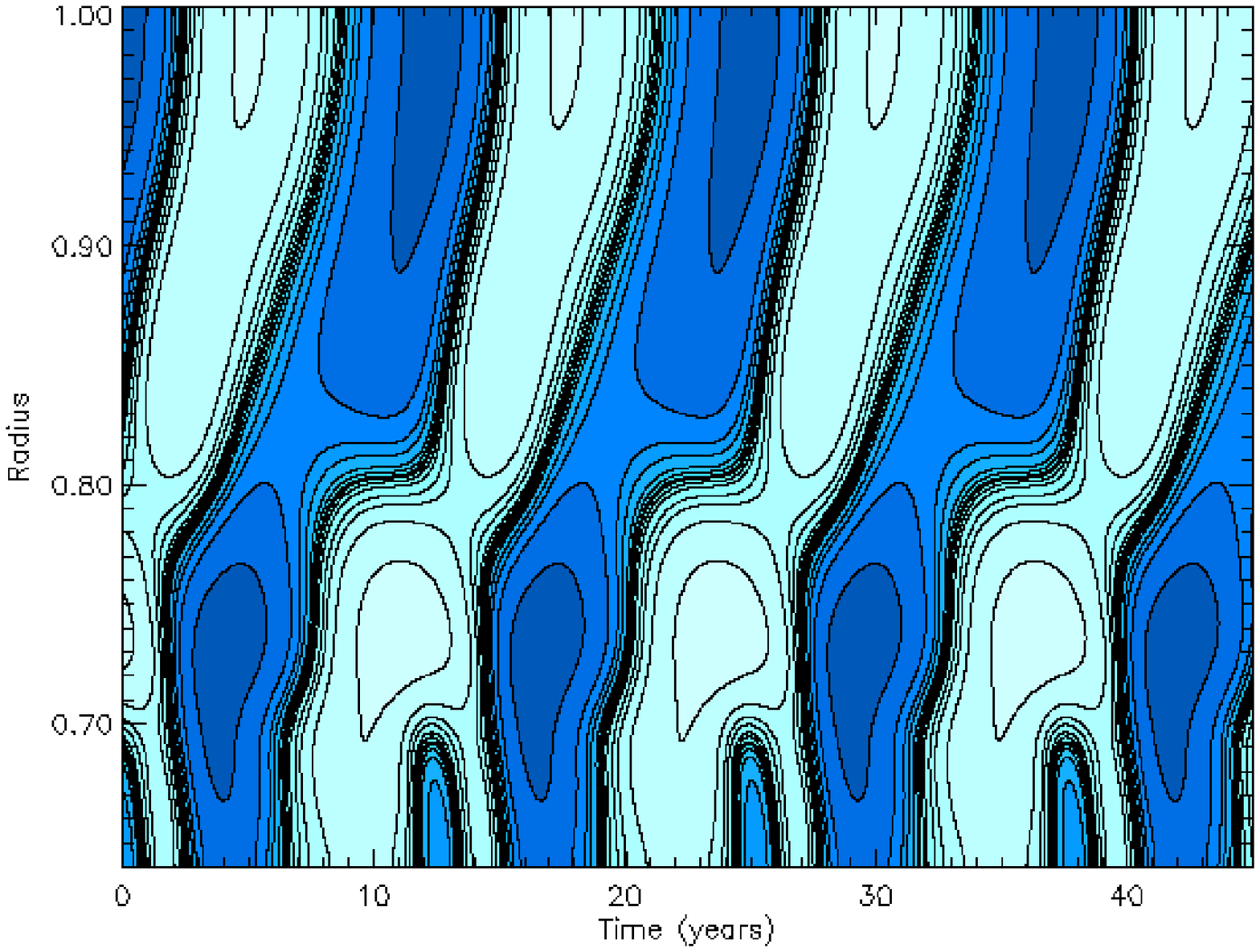}}
\end{center}
\caption{\label{fragmentation_VORONTSOV}
Radial contours of the angular velocity residuals $\delta \Omega$
as a function of time for a cut at latitude $30^{\circ}$,
with the zero order rotation profile obtained by
taking the first two (top panel) and the first three terms (bottom panel) in (\ref{expansion}).
Parameter values
are $R_\alpha=-20.0$, $P_r=1.0$ and ${R_{\omega}}=44000$.
}
\end{figure}

Both of the rotation profiles considered here produced
spatiotemporal fragmentation.
As examples, the top and the bottom panels of 
Fig.\ \ref{fragmentation_VORONTSOV} show 
the radial contours of the angular velocity residuals $\delta \Omega$
as a function of time for a cut at latitude $30^{\circ}$,
with the zero order rotation profile given by the first two and the first
three terms in the expression (\ref{expansion}), respectively.

To test further this robustness, we also used functions
$\Omega_{2i+1} (r), i=0,1,2$, obtained from an inversion with slightly
different regularisation parameters and in a somewhat
different frequency range of the input data (Vorontsov et al 2000).
We found no qualitative changes in the spatiotemporal fragmentation.

\begin{figure}[!htb]
\centerline{\def\epsfsize#1#2{0.4#1}\epsffile{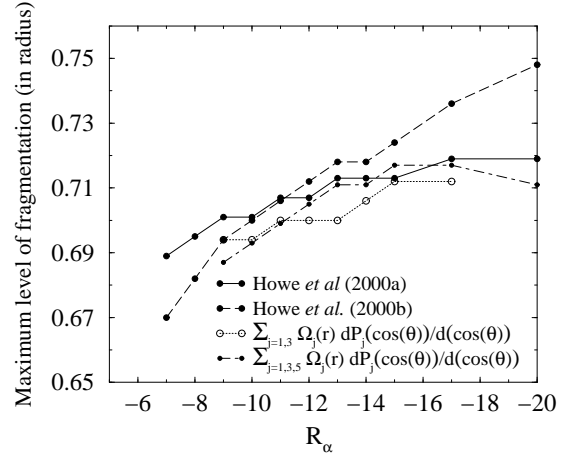}}
\caption{\label{fragmentation.MDI.GONG.Vorontsov.13.135} 
The height of the top of the fragmentation region for different zero order  rotation 
profiles $\Omega_0$ and different dynamo parameters $R_{\alpha}$.
}
\end{figure}

Finally, as a comparison with our previous results
obtained by employing the GONG and MDI data,
we have plotted in Fig.\ \ref{fragmentation.MDI.GONG.Vorontsov.13.135}
the radius of the top of the fragmentation region.
This qualitative similarity of the fragmentation in all the  cases 
is further evidence of robustness.

\section{Discussion}
We have shown that spatiotemporal fragmentation persists
even when only the lowest order terms in the Legendre expansion of the 
inversion for the solar rotation law are used. This is important, as these terms 
appear to be approximately time independent (over the scale of a solar cycle),
and also dominate the expansion. Indeed, these terms really represent the
rotation law that {\it should} be input to the dynamo code -- in principle solution
of Eq.\ (\ref{NS}) should then {\it determine} the variations of $\Omega$ over
a cycle. Thus, these results allay worries that the inconsistency
of averaging the observed rotation law over
only a part of a solar cycle might be giving misleading results.

The results presented here, together with those of CTM1 and CTM2,
show spatiotemporal fragmentation to be  a very persistent feature of
our simulations.
This finding suggests that nonlinear dynamo modelling has a real
potential predictive power, which could lead to a
better understanding of the solar dynamo when longer
data sets as well as improved analyses of them become available.
This in turn, in conjunction with the expected 
improvements in the helioseismological data,
could lead to an improved determination of the properties of the solar
interior, including the nature of the tachocline. 


\end{document}